\documentclass[twocolumn,showpacs,preprintnumbers,amsmath,amssymb]{revtex4}

\usepackage{graphicx}
\usepackage{dcolumn}
\usepackage{bm}
\usepackage{natbib}

\begin{document}

\preprint{to be published in JAP}

\title{Threshold Voltage Shift in Organic Field Effect Transistors by \\Dipole-Monolayers on the Gate Insulator}

\author{K.~P.~Pernstich}
    \email{Kurt.Pernstich@solid.phys.ethz.ch}
\author{A.~N.~Rashid}%
    \altaffiliation{Institute of Quantum Electronics, ETH Z\"{u}rich, CH - 8093 Z\"{u}rich, Switzerland}
\author{S.~Haas}
\author{G.~Schitter}
    \altaffiliation{Nanotechnology Group, ETH Z\"{u}rich, CH - 8093 Z\"{u}rich, Switzerland}
\author{D.~Oberhoff}
\author{C.~Goldmann}
\author{D.~J.~Gundlach}
\author{B.~Batlogg}
\affiliation{Laboratory for Solid State Physics, ETH Z\"{u}rich, CH - 8093 Z\"{u}rich, Switzerland}%

\date{\today}

\begin{abstract}
We demonstrate controllable shift of the threshold voltage and the
turn-on voltage in pentacene thin film transistors and rubrene
single crystal field effect transistors (FET) by the use of nine
organosilanes with different functional groups. Prior to
depositing the organic semiconductors, the organosilanes were
applied to the SiO$_2$ gate insulator from solution and form a
self assembled monolayer (SAM). The observed shift of the transfer
characteristics range from -2 to 50~V and can be related to the
surface potential of the layer next to the transistor channel.
Concomitantly the mobile charge carrier concentration at zero gate
bias reaches up to $4\times 10^{12}$/cm$^2$. In the single crystal
FETs the measured transfer characteristics are also shifted, while
essentially maintaining the high quality of the subthreshold
swing. The shift of the transfer characteristics is governed by
the built-in electric field of the SAM and can be explained using
a simple energy level diagram. In the thin film devices, the
subthreshold region is broadened, indicating that the SAM creates
additional trap states, whose density is estimated to be of order
$1\times 10^{12}$/cm$^2$.
\end{abstract}

\pacs{73.61.Ph 73.20.At 68.55.Jk 68.37.Yz 68.37.Ps}

\keywords{turn-on voltage, surface treatment, enhancement
mode operation, SAM-induced charge, surface potential}

\maketitle

\section{Introduction}
Organic semiconducting materials are used to fabricate
transistors with electronic properties comparable to a-Si:H
\cite{Klauk02}\cite{Kelley03}, a material often used for
back panel circuits of active matrix displays. These
comparable electronic characteristics together with the
promising low-cost fabrication \cite{Moore02} makes organic
materials attractive candidates for use in commercial
products. However, to manufacture integrated circuits with
organic transistors the precise control of all electrical
properties is required. In addition to the charge carrier
mobility, the threshold voltage ($V_t$) is an important
parameter that needs to be controlled to ensure proper
operation of the circuits. The threshold voltage can depend
on the time a gate voltage has been applied (bias stress)
\cite{Knipp03}\cite{Salleo03}\cite{Street03}\cite{Northrup03}\cite{Gomes04},
on the exposure of the device to light \cite{Voelkel02} or
it can be shifted using a polarizable gate insulator
\cite{Katz02}. Furthermore, a dependence on the work
function of the gate electrode \cite{Li02} and the
thickness of the active layer material \cite{Schroeder03}
has been reported. As we will show in this article, the
threshold voltage additionally depends strongly on the
preparation of the surface on which the organic material is
deposited.

We present an experimental method to systematically study
the influence of the surface treatment of the gate
insulator on the threshold voltage and other electrical
properties of pentacene thin film transistors (TFTs) and
rubrene single crystal FETs. Top contact pentacene TFTs
were fabricated on heavily doped and oxidized silicon
wafers. Prior to the pentacene deposition the silicon
dioxide gate insulator was treated with solutions of a
variety of organosilanes with different degrees of electron
acceptance properties. The organosilanes form self
assembled monolayers on the SiO$_2$ gate insulator and can
advantageously modify the electronic properties of thin
film transistors \cite{Lin97}\cite{Salleo02}. The single
crystal FETs were fabricated by placing freshly grown
crystals onto prepatterned wafers covered with various
SAMs. The SAMs have a built-in dipole field depending on
the molecule's functional group and modify the (mobile)
charge carrier density. This SAM-induced modification of
the charge carrier density in the transistor channel is
similar to applying a gate voltage. Both, the threshold
voltage and the turn-on voltage are governed by the
built-in electric field of the SAM. Similar results for
bottom contact transistors have been reported recently by
Kobayashi et al. \cite{Kobayashi04}.

The transfer characteristics of the single crystal devices
are shifted by a certain gate voltage depending on the SAM,
while maintaining a steep subthreshold swing. The thin film
devices however, show a pronounced broadening of the
subthreshold region. From this broadening an increased trap
density is extracted that can partly be explained by a poor
film morphology as observed with X-Ray diffraction (XRD)
and atomic force microscopy (AFM) measurements, and partly
by additional trap states.

\begin{figure}
\includegraphics{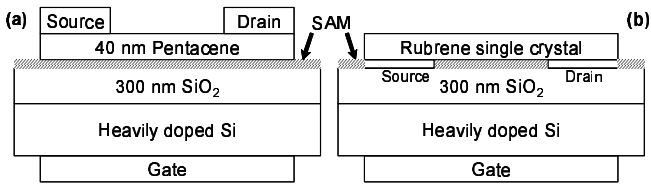}
\caption{\label{fig:device}Schematic device structure of the
inverted-staggered pentacene thin film transistors (a) and the
rubrene single crystal FETs (b). The molecules used for the self
assembled monolayer (SAM) are shown in Fig. \ref{fig:molecules}.}
\end{figure}
\section{Experimental}\label{sec:experimental}
Figure \ref{fig:device} shows a schematic device cross section for
thin film transistors (panel a) and single crystal FETs (panel b).
Heavily doped silicon wafers with a 300~nm thick silicon dioxide
insulating layer were used as substrates. The wafers were
successively cleaned in hot acetone and hot isopropanol for three
minutes in an ultrasonic bath, then with a piranha solution
(70~vol\%~$H_2SO_4$~:~30~vol\%~$H_2O_2\cdot 30\%$) for approx. 20
minutes, and were finally thoroughly rinsed in ultrapure water.
The substrates were treated in a glove box with a relative
humidity near 3~\%. The treatment process was optimized for
octadecyltrichlorosilane (OTS) and was applied in the same way for
the other organosilanes.

\begin{figure}
\includegraphics{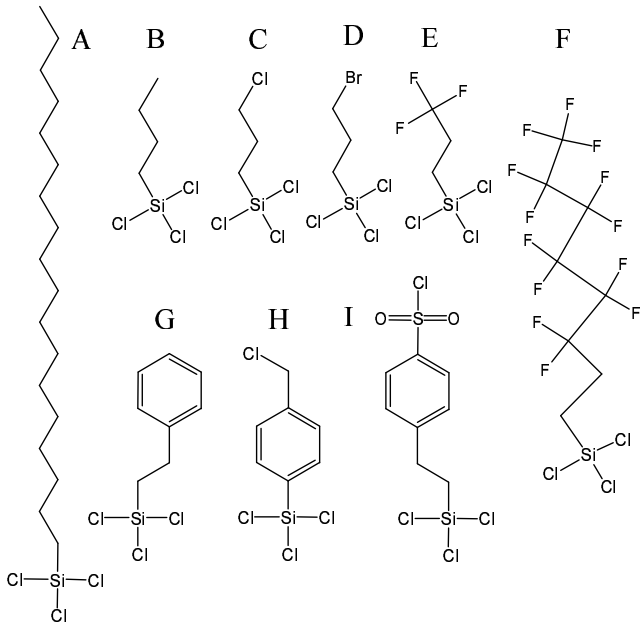}
\caption{\label{fig:molecules}Molecular structure of the studied
organosilanes; carbon and hydrogen atoms are not shown. In
monolayer formation, the Cl-atoms from the anchor group (SiCl$_3$)
are removed and the Si bonds covalently to the SiO$_2$ surface as
well as to neighboring Si-atoms.}
\end{figure}
To form the SAM, the wafers were immersed for 3~h in a 3~mM
solution of the organosilane in anhydrous toluene
\cite{McGovern94}. Fig. \ref{fig:molecules} shows the molecular
structures of the studied organosilanes. After removing the
samples from the solution they were cleaned in fresh toluene for 2
minutes in an ultrasonic bath to remove any excessive layers
\cite{Tillman88}. We found this step to be crucial for good
monolayer formation. The monolayers were then baked on a hot plate
for 1~h at 150~$^\circ$C \cite{Angst91} in the same glove box to
enhance cross-linking of the organosilane molecules and covalent
bond formation to the silica surface.

To fabricate TFTs, the samples were transferred into the
deposition chamber where pentacene was deposited at a rate
of 0.3~$\pm$~ 0.1~\AA/sec, by thermally evaporating
pentacene powder that had previously been purified twice by
temperature gradient vacuum sublimation. The nominal
thickness of the organic layer was 40~nm and the base
pressure of the system was near $1\times 10^{-6}$~mbar.
Unless otherwise noted, the substrate temperature during
deposition was kept at 50~$^\circ$C. In every deposition
batch we deposited pentacene onto eight wafers at a time.
Two of those wafers were treated with OTS to check the
quality of the fabrication process and the rest were
treated with three other organosilanes.

Gold source and drain contacts were deposited through
shadow masks at a rate near 1~\AA/sec. The channel width W
was 600~$\mu$m for all devices while the gate length L
varied from 30~$\mu$m to 150~$\mu$m. With this
configuration we could fabricate six transistors on every
wafer. The electrical properties were measured with a
HP~4155A semiconductor parameter analyzer, with the samples
kept in an argon glove box ($<0.1$~ppm H$_2$O, O$_2$).

For the single crystal experiments, 20~nm thick gold source
and drain contacts were evaporated after cleaning the
wafers, forming bottom contacts. After finishing the
treatment process, rubrene single crystals grown as
described in \cite{Goldmann04} were carefully placed on the
prepatterned structures completing the transistors. The
measurement procedure was the same as for the thin film
transistors.

\section{Results and Discussion}
\begin{figure}
\includegraphics{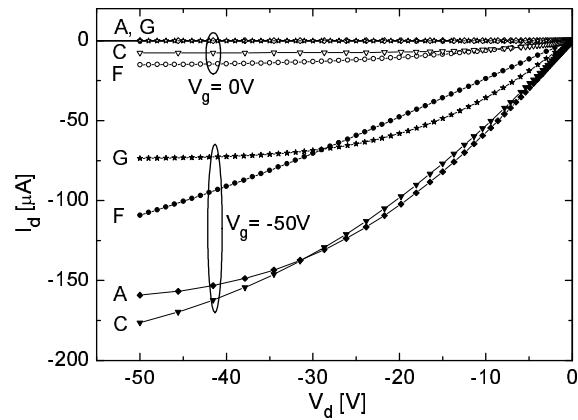}
\caption{\label{fig:output} Output characteristics of TFTs with
four different surface treatments (A)(C)(F)(G). At zero gate bias
the drain current of (C) and (F) is non-zero, indicating the
presence of mobile charge carriers even without gate bias. The
drain current of (F) does not saturate at $V_g=-50$~V indicating
that the transistor is still operating in the linear regime
($V_d<V_g-V_t$). The transistor geometry is $W=600~\mu m$ and
$L=30~\mu m$. The pentacene was deposited at 50~$^\circ$C.}
\end{figure}

The output characteristics of TFTs with four different surface
treatments are shown in Fig. \ref{fig:output}. A non-zero drain
current at zero gate bias is measured for transistors (C) and (F)
while it is zero on a linear scale for the OTS treated transistor
(A) and transistor (G). This indicates the presence of mobile
charge carriers at zero gate bias. The drain current at negative
gate bias saturates in transistors with treatments (A) and (G),
following the standard MOSFET behavior \cite{Sze}. For transistor
(C) this saturation is not as pronounced and for transistor (F) no
saturation is observed, indicating a large positive threshold
voltage (V$_t$) so that the device is still operating in the
linear regime ($V_d<V_g-V_t$).

\begin{figure}
\includegraphics{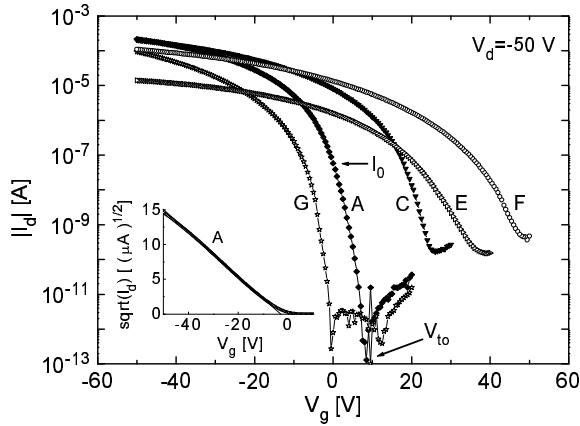}
\caption{\label{fig:transfer} Transfer characteristics of the same
transistors as in Fig. \ref{fig:output} and the transistor with
treatment (E). The turn-on voltage shifts towards more positive
values for treatments (G)(A)(C)(E) and (F). The increase of $I_0$
indicates the presence of mobile charge carriers at zero gate
bias. The values of $I_0$ and $V_{to}$ are marked for treatment
(A). The inset shows the extraction of V$_t$ and $\mu$.}
\end{figure}
To illustrate the presence of mobile charge carriers at zero gate
bias we show in Fig. \ref{fig:transfer} the transfer
characteristics as log$|I_d|$ versus $V_g$ for the same
transistors as in Fig. \ref{fig:output} and the transistor with
treatment (E). The magnitude of the drain current $I_0$ at
$V_g=0$~V shows a dependence on the SAM and is marked in Fig.
\ref{fig:transfer} for the OTS (A) treated transistor. In addition
to the increase of $I_0$, an increased off-current is observed for
treatments (C)(E) and (F).

To quantify the measurements we use the turn-on voltage
$V_{to}$, which is the gate voltage where the drain current
starts to increase exponentially. For polymer devices a
switch-on voltage was defined in a similar way
\cite{Meijer02} and marks the flat band condition. The
turn-on voltage is marked in Fig. \ref{fig:transfer} for
transistor (A). It is slightly positive for (A)
$V_{to,A}=8.5$~V which is commonly observed for OTS treated
devices, and it increases significantly for treatment (C)
$V_{to,C}=25$~V, (E) $V_{to,E}=36$~V and (F)
$V_{to,F}=49$~V. The only treatment with negative turn-on
voltage is the phenyl treatment (G) $V_{to,G}=-1.5$~V,
meaning that the transistor is completely switched off at
zero gate bias and the transistor is operating entirely as
an enhancement mode device, which can be desirable for
designing circuits.

\begin{table*}
\caption{\label{tab:table}Summarized properties resulting from the
different surface modifications. $\Theta$ is the average contact
angle of water with the surface measured on two different
substrates. $\mu$ is the charge carrier mobility, $V_t$ the
threshold voltage and $V_{to}$ the turn-on voltage of the TFTs.
$S$ is the subthreshold swing (300~nm SiO$_2$) and $I_0$ is the
drain current at zero gate bias. The given values represent the
mean value (standard deviation) over typically 9 transistors
fabricated on two different substrates in the same batch.}
\begin{ruledtabular}
\begin{tabular}{l|c|c|c|c|c|c|c}
 & $\Theta$ & $\mu$ & $V_t$ & $V_{to}$ & $S$ & $|I_0|$\\
& [$^\circ$] & [cm$^2$/Vs] & [V] & [V] & [V/dec.] & [A]\\
\hline
(A) Octadecyltrichlorosilane & 95 & $0.96 (16)$ & $-3.7(1.0)$ & $4.7$ & $0.9$ & $10^{-8}$\\
(B) Butyltrichlorosilane & 93 & $0.61 (11)$ & $-4.3(0.5)$ & $4.7$ & $1.1$ & $10^{-8}$\\
(C) 3-Chloropropyltrichlorosilane & 75 & $0.71 (09)$ & $1.5(1.8)$ & $16$ & $1.8$ & $10^{-6}$\\
(D) 3-Bromopropyltrichlorosilane & 80 & $0.74 (13)$ & $2.8(2.8)$ & $17$ & $2$ & $10^{-6}$\\
(E) Trichloro(3,3,3-trifluoropropyl)silane & 91 & $0.03 (01)$ & $22.7(5.2)$ & $33$ & $4.9$ & $10^{-7}$\\
(F) 1H,1H,2H,2H-Perfluorooctyl-trichlorosilane & 105 & $0.15 (02)$ & $26(2.0)$ & $44$ & $4.9$ & $10^{-6}$\\
(G) Phenethyltrichlorosilane & 92 & $0.71 (11)$ & $-12.7(1.2)$ & $-1.5$ & $0.9$ & $10^{-12}$\\
(H) 4-(Chloromethyl)phenyltrichlorosilane & 88 & $0.56 (12)$ & $-7(1)$ & $4$ & $1.2$ & $10^{-8}$\\
(I) 2-(4-Chlorosulfonylphenyl)ethyltrichlorosilane & 90 & $0.36 (05)$ & $25(3)$ & $49$ & $4.4$ & $10^{-5}$\\
\end{tabular}
\end{ruledtabular}
\end{table*}
Table \ref{tab:table} summarizes the results for $V_t$, $V_{to}$,
$I_0$, subthreshold swing S (for the 300~nm thick SiO$_2$ gate
insulator) and calculated mobility $\mu$ for the nine treatments,
together with the water contact angle $\Theta$ of the treated
surface. The threshold voltage was defined as the intercept of a
linear least square fit to $\sqrt I_d$ versus $V_g$ as illustrated
in the inset to Fig. \ref{fig:transfer}. The range between 20\%
and 80\% of $I_{d, max}$ was taken for this fit, and the mobility
was calculated from the slope. The subthreshold swing was
extracted from the logarithmic plot of the transfer
characteristics shown in Fig. \ref{fig:transfer}. The values in
Tab. \ref{tab:table} represent the average values and the standard
deviation measured on typically nine transistors fabricated on two
different wafers in the same batch.

We will discuss two mechanisms possibly involved in the
shift of the threshold voltage and the turn-on voltage: the
influence of the film morphology and the effect of the
built-in electric field of the SAM ("SAM-induced charge").

\subsection{Influence of the Film Morphology}

\begin{table}
\caption{\label{tab:tabletemp}Charge carrier mobility $\mu$,
threshold voltage $V_t$, turn-on voltage $V_{to}$ and estimated
trap density $N_{trap}$ of transistors with pentacene films
deposited at three different substrate temperatures. The gate
insulator was treated with (A), (B) and (I). Except the mobility
no parameter is significantly affected by the different deposition
temperatures hence the film morphology, demonstrating the dominant
effect to be the treatment with the different organosilanes. The
trap density is estimated from the threshold voltage above turn-on
voltage (see text for details).}
\begin{ruledtabular}
\begin{tabular}{c|c|c|c|c|c}
 & T [$^\circ$C] & $\mu$~[cm$^2$/Vs] & $V_t$~[V] & $V_{to}$~[V] & $N_{trap}$~[$10^{12}$/cm$^2$]\\
\hline
 & 30 & 0.4(1) & -8(3) & 2(5) & 0.7(6)\\
(A) & 50 & 0.9(1) & -4(1) & 4(2) & 0.6(2)\\
 & 70 & 1.3(2) & -10(2) & 0(3) & 0.7(4)\\
 \hline
 & 30 & 0.6(1) & -11(1) & -1(1) & 0.9(1)\\
(B)& 50 & 0.7(1) & -4(0.5) & 5(2) & 0.6(2)\\
 & 70 & 0.9(1) & -11(1) & 0(0.5) & 0.8(1)\\
 \hline
 & 30 & 0.4(.05) & 24(6) & 45(8) & 1.5(1.0)\\
(I)& 50 & 0.4(.05) & 25(3) & 48(2) & 1.6(4)\\
 & 70 & 0.3(.02) & 26(2) & 50(1) & 1.7(2)\\
\end{tabular}
\end{ruledtabular}
\end{table}
The film morphology has been shown to influence the charge carrier
mobility
\cite{Knipp03}\cite{Lin97}\cite{Shtein02}\cite{Gundlach03}.
Especially the morphology of the first few monolayers where charge
transport occurs \cite{Voelkel02}\cite{Tanase03} is expected to
strongly influence the mobility \cite{Gundlach03}\cite{Dinelli04}.
To investigate the influence of the film morphology on the
threshold and the turn-on voltage, we fabricated transistors with
treatments (A), (B) and (I) where the pentacene had been deposited
at 30, 50 and 70~$^\circ$C. Only a weak dependence and no general
trend was observed between the film morphology as characterized by
AFM measurements and X-Ray diffraction (XRD) and the threshold and
turn-on voltage. Listed in Tab. \ref{tab:tabletemp} are the
mobility, threshold voltage and turn-on voltage for those
transistors, as well as an estimated trap density which is
discussed below.

\begin{figure}
\includegraphics{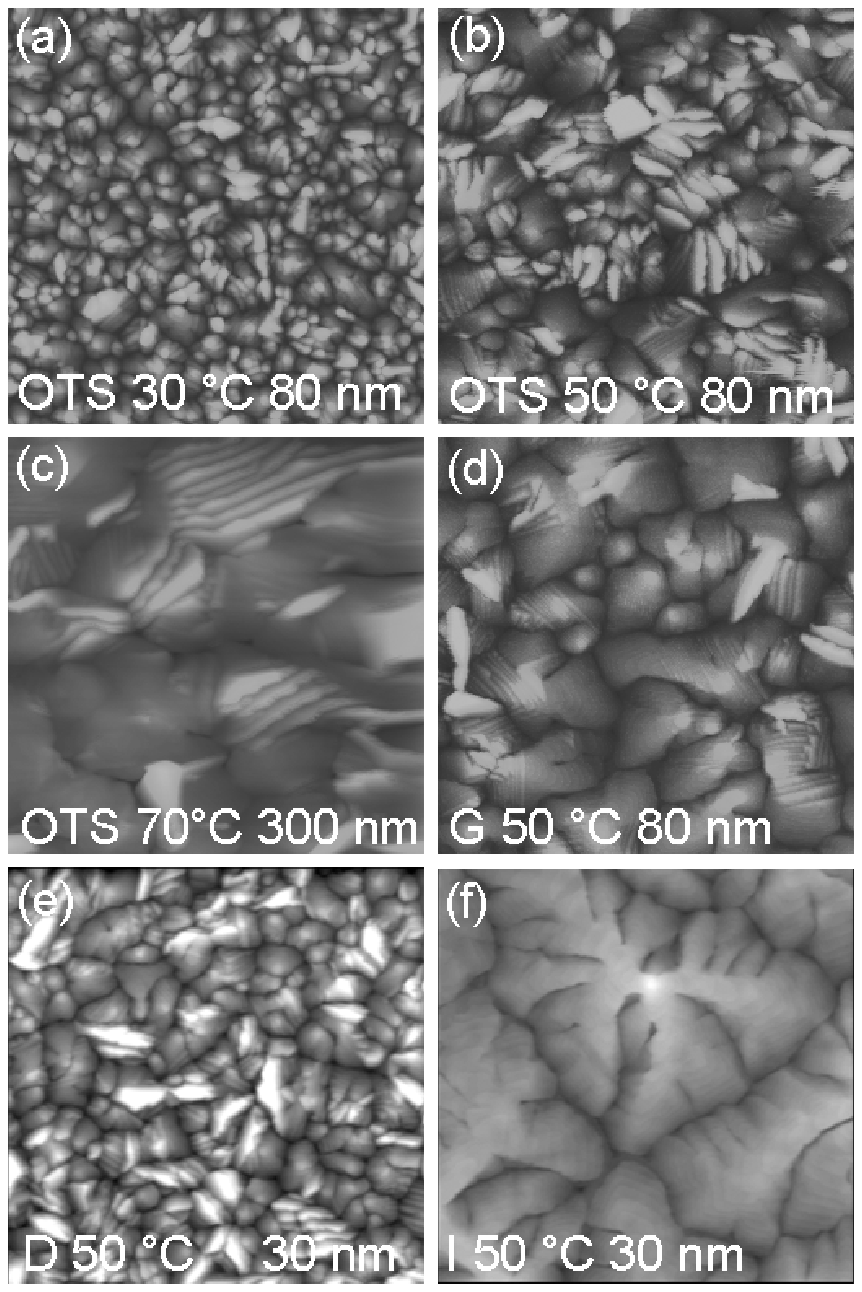}
\caption{\label{fig:AFM}Topographical images of pentacene films
deposited at various substrate temperatures and substrate
treatments as indicated in the images. The scan area is
5x5~$\mu$m$^2$ for all images. The grain size increases with
temperature (a)-(c), and the brightest spots mark grains that
presumably consist of flat lying molecules. For the phenyl
treatment (d) the grain size increases with respect to the OTS
counterpart (b), simultaneously decreasing the density of edge
oriented grains. Although the grain size in (e) is smaller than in
(f) the mobility of transistors fabricated on (D) is larger than
on (I).}
\end{figure}
Pentacene films deposited at higher substrate temperature often
consist of large grains \cite{Hajlaoui02}. The larger grain size
forming at higher temperatures on OTS treated substrates is shown
in the topography images in Fig. \ref{fig:AFM}(a)-(c). The images
show evidence of lamellar growth and the brightest spots mark
grains with a height well above the average film thickness. Those
grains consist presumably of flat lying pentacene molecules
\cite{Shtein02}\cite{Gundlach03}. The height of those grains
increases with increasing deposition temperature, indicating a
rapid growth of grains in the a-b plane. A typical image of a film
deposited onto a phenyl treated (G) substrate is shown in Fig.
\ref{fig:AFM}(d) and reveals larger grains than obtained for films
on OTS treated substrates held at the same deposition temperature
(cf. Fig. \ref{fig:AFM}(b)). Additionally, the density of edge
oriented grains (resulting from flat lying molecules) is smaller
for films deposited onto substrates treated with (G). In Fig.
\ref{fig:AFM}(e) and \ref{fig:AFM}(f) topography images of films
deposited onto substrates treated with (D) and (I) show an
opposite relationship between the mobility and the grain size:
although the grain size of film (D) is smaller compared to (I),
the transistors on (D) show a larger mobility than the one on (I)
(cf. Tab. \ref{tab:table}). It is worthwhile emphasizing that the
film morphology as observed with AFM does not necessarily reflect
the microstructure of the first few monolayers of pentacene that
forms the electrically active channel.

\begin{figure}
\includegraphics{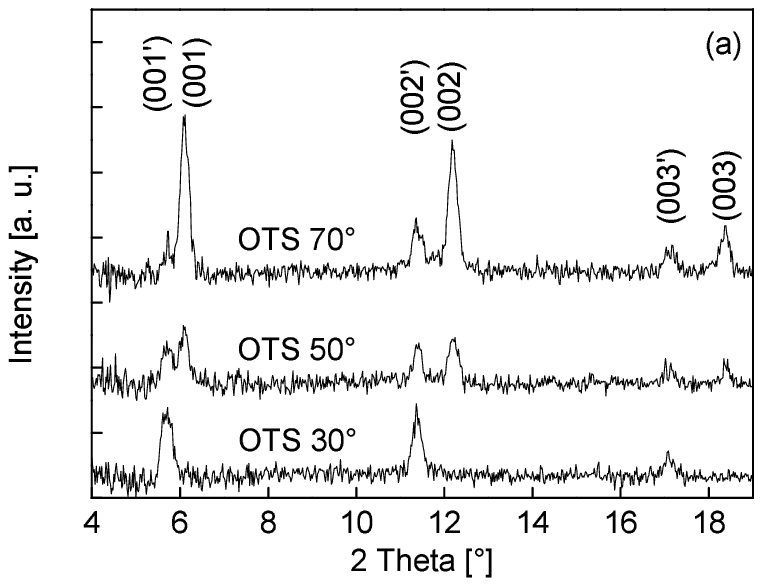}
\includegraphics{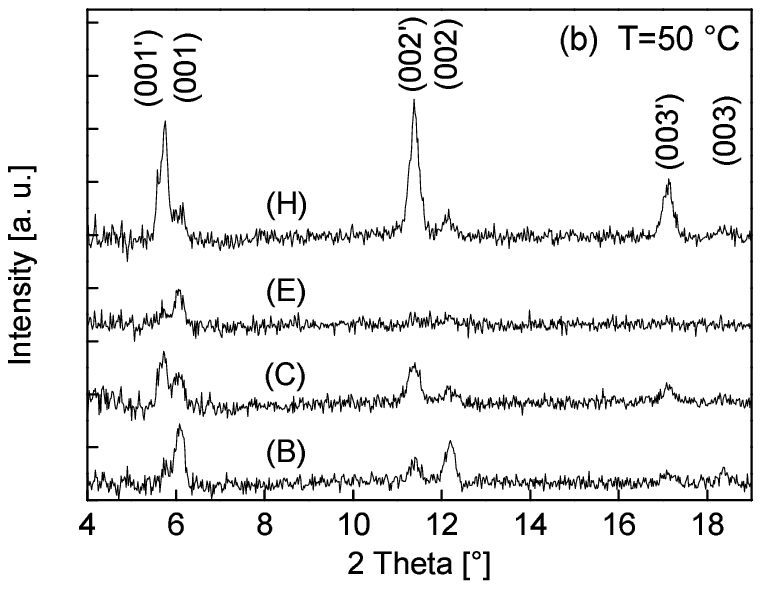}
\caption{\label{fig:XRD}X-Ray diffraction pattern of 40~nm thick
pentacene films deposited onto OTS treated substrates at 30, 50
and 70~$^\circ$C (a), and on substrates treated with (B), (C), (E)
and (H) held at 50~$^\circ$C (b). The "single crystal phase"
becomes more dominant with higher deposition temperatures (a). The
abundance of the "thin film phase" and "single crystal phase"
varies with the surface treatment (b), indicating differences in
over-all film morphology.}
\end{figure}
The XRD patterns in Fig. \ref{fig:XRD} reveal the over-all
difference in film morphology: pentacene films generally show two
distinct crystalline phases with different d-spacings, a "thin
film phase" with 15.4~\AA\ (00$\ell'$) and a "single crystal
phase" with 14.4~\AA\ (00$\ell$)
\cite{Dimitrakopoulos96}\cite{Bouchoms99}\cite{Mattheus03}. The
films deposited at 30~$^\circ$C crystallize in the "thin film
phase" only, while the "single crystal phase" is more prevalent in
the films deposited at higher substrate temperatures \cite{Lin97b}
(cf. Fig. \ref{fig:XRD}(a)). For the OTS devices, the hole
mobility is slightly higher in the latter as can be seen from Tab.
\ref{tab:tabletemp}. This is in agreement with results reported in
e.g. \cite{Knipp03}, and is presumably due to better overlap of
the $\pi$-orbitals \cite{Bredas02} of the pentacene molecules. In
Fig. \ref{fig:XRD}(b) the mixture of the "thin film phase" and the
"single crystal phase" is shown for treatments (B), (C), (E) and
(H). Here the pentacene films have been deposited on substrates
held at 50~$^\circ$C and interestingly, a different trend is
observed for treatments (C) and (B): TFTs on SAM (B) show a lower
mobility although the single crystal phase is more dominant. From
a comparison of the results in Tab. \ref{tab:tabletemp} we
conclude that the variations of the TFT characteristics ($\Delta
V_t$, $\Delta V_{to}$) are dominated by the particular
organosilanes forming the SAM and not by the over-all film
morphology as probed by XRD and AFM. This conclusion is supported
by the single crystal experiments described later.
\\

%
\subsection{Effect of the SAM's dipole field}
The observed shifts in the electrical characteristics
correspond to the electron acceptance properties of the
organosilane molecule's end group. For treatment (C) with
the CH$_2$Cl end group for instance, this means that
electrons from the pentacene film are attracted by the SAM
leaving behind mobile holes in the channel. Thus a more
positive gate bias is needed to switch off the device, i.e.
$V_{to}$ shifts towards more positive values.

The electronegativity of the molecule's functional group
influences the charge distribution within the molecule and can
lead to the formation of an electric dipole. Campbell et al.
\cite{Campbell96} calculated the charge distribution within
similar molecules using an ab initio scheme and found a dipole
moment whose strength depends on the functional group of the
investigated molecule. When such molecules form a SAM the
molecular dipoles gives rise to a net polarization of the SAM that
changes the surface potential \cite{Ishii99} as verified with
Kelvin-probe measurements in \cite{Campbell96} and by Kelvin-probe
force microscopy in \cite{Sugimura02}. In \cite{Sugimura02} the
authors calculated dipole moments of 0.5 and -1~Debye for isolated
molecules similar to (A) and (F), and measured a surface potential
difference of approximately 0.2~V between the corresponding SAMs
formed on SiO$_2$. Assuming the thickness of the SAMs to be 2~nm,
this corresponds to an electric field of 1~MV/cm. To produce the
same field by applying a voltage across the 300~nm thick SiO$_2$
gate insulator a gate voltage of 30~V is necessary, which
corresponds well with the shifts in transfer characteristics we
measure in our devices (cf. Tab. \ref{tab:table}).

In the presented situation the charge density respectively
the energy levels need to be considered in a
self-consistent way, resulting from the properties of the
individual molecule in the SAM attached to silica, and the
adjacent pentacene molecules. This is important because it
has been shown in e.g. \cite{Vager02} that the electronic
properties of a close-packed organized organic monolayer
can differ from the properties of the isolated molecule.
Additionally our samples were exposed to ambient air where
water can adsorb on the surface which might affect the
effective dipole strength of the SAM.

\begin{figure}
\includegraphics{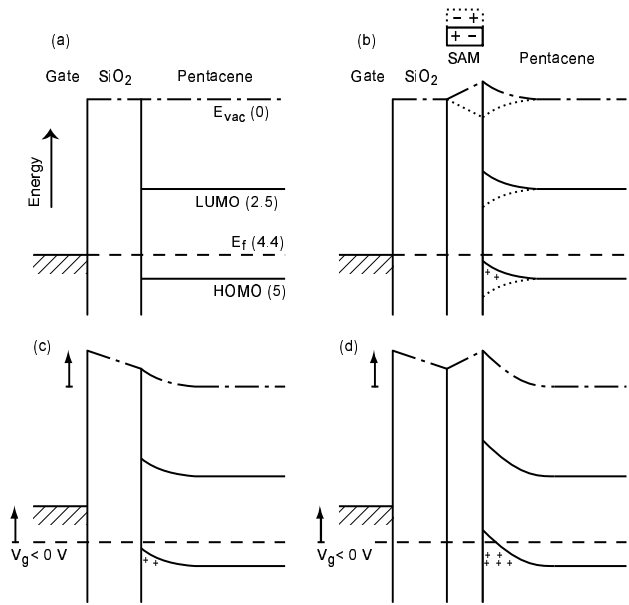}
\caption{\label{fig:Bands}Schematic energy level diagram suggested
for surface treated TFTs. For an untreated SiO$_2$ surface the
vacuum levels of the SiO$_2$ and the pentacene are aligned and no
band bending occurs (a). In (c) a negative gate voltage is applied
shifting the gate electrode's Fermi level towards higher energies
and bending the HOMO and LUMO levels of the pentacene. In (b) the
permanent dipole field of the SAM shifts the surface potential
which has the same effect as applying a gate voltage. In (d) a
combination of (b) and (c) is shown. The given numbers are taken
from reference \cite{Watkins03} and are given in $eV$.}
\end{figure}
The change in surface potential modifies the interface properties
as illustrated in the schematic band diagram shown in Fig.
\ref{fig:Bands}. When pentacene is deposited onto SiO$_2$ under
UHV conditions, the vacuum levels are aligned and no bending of
the highest occupied molecular orbital (HOMO) and lowest
unoccupied molecular orbital (LUMO) level occurs \cite{Watkins03}
as illustrated in Fig. \ref{fig:Bands}(a). For simplicity, only
the gate electrode's Fermi level is shown. When a negative gate
voltage is applied the Fermi level of the gate electrode shifts
towards higher (electron) energies. Part of the applied gate
voltage is dropped across the gate insulator, and since the band
alignment of the HOMO and LUMO level is fixed with respect to the
vacuum level, the remaining gate voltage bends the HOMO and the
LUMO levels. Therefore mobile charge carriers can accumulate and
form the conducting channel. For a SAM with a permanent electric
dipole field inserted between the gate insulator and the
pentacene, the situation is as illustrated in Fig.
\ref{fig:Bands}(b): the dipole field of the SAM modifies the
surface potential which has the same effect as applying a
(negative) gate voltage. The solid curves in Fig.
\ref{fig:Bands}(b) appear to be valid for all treatments except
the phenyl treatment (G). For this treatment the situation may be
depicted with the dotted curves where the majority carriers are
depleted. Fig. \ref{fig:Bands}(c, d) depict the situation where a
negative gate voltage is applied to devices with (Fig.
\ref{fig:Bands}(d)) and without (Fig. \ref{fig:Bands}(c)) a SAM:
the gate voltage rises the vacuum level of the gate insulator and
additionally it is raised by the permanent dipole field of the
SAM, resulting in an increased band bending and therefore in an
increased hole density in the channel. As a consequence, the
turn-on and the threshold voltage are determined by the surface
potential of the layer next to the transistor channel. We
emphasize that any surface charge present at the gate insulator
due to a contact potential \cite{Tsividis99} or imperfections such
as oxygen (OH-groups), water molecules\cite{Nicollian91} or mobile
ions \cite{Rep03} also influences the surface potential and
therefore influences the threshold voltage and the turn-on
voltage; especially in devices with untreated oxide
\cite{Gundlach03}. Thus Fig. \ref{fig:Bands} may capture only part
of the total situation relevant for the device performance.
\\

\begin{figure}
\includegraphics{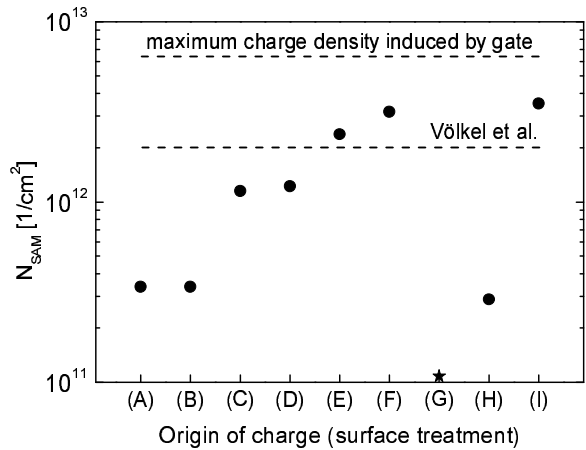}
\caption{\label{fig:Nsam}Induced charge carrier density by the
different SAMs on the gate insulator. Round symbols represent hole
accumulation and the star represents hole depletion. The maximum
charge carrier density induced by the gate was obtained using a
gate field of 3~MV/cm. The value marked as "V\"{o}lkel et al." is
deduced from simulations in \cite{Voelkel02}.}
\end{figure}
The mobile charge carrier density induced by the SAM ($N_{SAM}$)
is shown for the various SAMs in Fig. \ref{fig:Nsam}. The density
was estimated using $N_{SAM}=C_{ox}\cdot V_{to}/e$, with
$C_{ox}=11.5$~nF/cm$^2$ being the measured insulator capacitance
per unit area and $e$ the elementary charge. The turn-on voltage
is chosen as it is a measure of the hole concentration in the
channel at zero gate bias: applying the turn-on voltage to the
gate electrode depletes the channel and the bulk pentacene as much
as possible. Using the flat band voltage would give a more
accurate estimate but it is not accessible from our measurements.
Since the pentacene films are assumed to be thinner than the
screening length near flat band condition \cite{Greve98}, we
expect that the turn-on voltage is very close to the flat band
voltage. The maximum carrier concentration corresponds to about
one induced mobile hole per 100 SAM molecules, assuming a surface
density of the SAM molecules of $4\times 10^{14}$~/cm$^2$
\cite{Vuillaume96}.
\\

%
\subsection{Density of States, Threshold Voltage and Additional Trap
States} A more microscopic approach taking into account the
imperfections of the semiconductor is desirable. Following
V\"{o}lkel et al. \cite{Voelkel02} the mobile holes observed at
zero gate bias can be modeled using electron acceptor
states in the band gap close to the HOMO level. Such
acceptor states move the Fermi level closer to the HOMO
level by changing the thermodynamic equilibrium position of
the Fermi level close to the gate insulator. V\"{o}lkel and
co-workers \cite{Voelkel02} used a one-dimensional
transistor model to study the effects of localized band-gap
states on the electrical characteristics of pentacene TFTs.
The authors introduced acceptor states at the interface
layer next to the gate insulator in order to explain their
observed shifts in turn-on voltage, and donor states to
account for the shifts in threshold voltage. A total trap
density of $4.8\times 10^{18}$/cm$^3$ accounts for the
observed shifts. Assuming a channel thickness of 5~nm and a
homogeneous carrier density in the channel, the acceptor
concentration in their model corresponds to a surface
charge density of $2.4\times 10^{12}$/cm$^2$, close to our
results given in Fig. \ref{fig:Nsam}.
\\

With increasing negative gate voltage, more trap states are
filled. If all deep traps are filled and the local Fermi energy in
the channel is in the energy range of the transport level (the
energy at which thermal activation begins to predominate)
\cite{Monroe85}\cite{Arkhipov03} the threshold voltage is reached
\cite{Shur84}\cite{Lang03}. Horowitz and Delannoy expressed this
condition as the equilibrium between trapped and mobile carriers
\cite{Horowitz91} (for a refinement see \cite{Balakrishnan03}).
Therefore the threshold voltage is tied to the turn-on voltage via
the trap density, and the threshold voltage above turn-on voltage
($V_{tto}=V_t-V_{to}$) is an estimate of the trap density in the
channel.

Estimating the number of trap states from the threshold
voltage above turn-on voltage we find the total trap
density $N_{trap}$ using $N_{trap}=C_{ox}\cdot |V_{tto}|/e$
where $C_{ox}$ is again the oxide capacitance per unit
area. This results in a trap density of $0.5-2\times
10^{12}$/cm$^2$. Taking into account only transistors with
mobilities greater than 0.5~cm$^2$/Vs the trap density in
the channel is estimated to be $0.5-1\times
10^{12}$/cm$^2$. The values are in good agreement with
values derived from simulations reported in
\cite{Voelkel02}.

The origin of the increased trap density cannot clearly be
revealed by these experiments. Our control experiments
suggest that the effect of the SAM dominates over the
effect of film morphology. Increased trap densities were
also found in polymer devices with a high-k gate insulator
compared to low-k gate insulators and were ascribed to a
dipolar disorder caused broadening of the Gaussian
distributed transport states \cite{Veres03}. In
\cite{Kadashchuk93} the authors report on dipole impurities
in anthracene single crystals and suggest that traps are
formed as a result of the interaction of carriers with the
dipole moment of the impurities. Similarly, the
introduction of dipole moments between gate insulator and
pentacene might change the local polarization of individual
pentacene molecules, therefore introducing new trap states
\cite{Silinsh94}.
\\

\subsection{Single Crystal Experiments}
To verify that the dipole field of the SAM governs the
turn-on voltage, and to test whether or not the strong
dipole moment of the SAM molecules can influence the trap
distribution of single crystal FETs (SC-FETs) we fabricated
SC-FETs using the "flip-crystal" technique
\cite{Takeya03}\cite{deBoer03}. Rubrene crystals were used
because large planar crystals can be grown as described in
e.g. \cite{Goldmann04}, and because rubrene shows a very
high mobility
\cite{Williams71}\cite{Goldmann04}\cite{Sundar04}. The
wafers were treated with (A) or (E) and the resulting
transistors have a on/off ratios $>10^7$ and a mobility
$>1$~cm$^2$/Vs, indicating that the presence of the SAM has
little or no influence on the effective mobility. It also
suggests that the low mobility of TFTs with treatment (E)
is probably caused by a poor molecular ordering as revealed
by XRD measurements.

\begin{figure}
\includegraphics{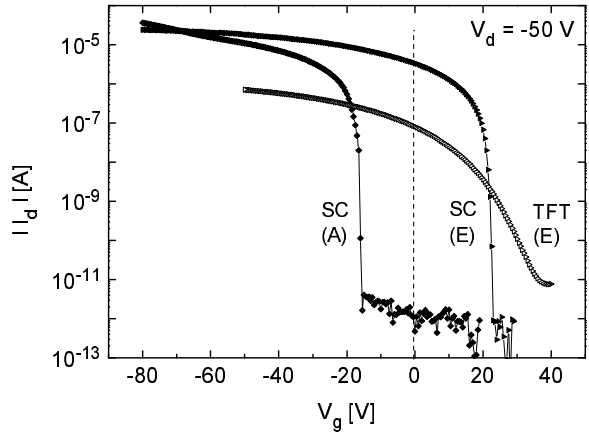}
\caption{\label{fig:FCTransfer}Transfer characteristics of rubrene
single crystal FETs treated with (A) and (E) and of a typical
pentacene thin film device treated with (E). The turn-on voltage
shifts towards more positive gate voltages for treatment (E). For
the single crystal devices the off-current is lower and the
subthreshold swing is steeper than for the thin film device. The
curve for the thin film transistor was shifted to correspond to
the same W/L ratio as in the single crystal FETs.}
\end{figure}
The transfer characteristics of the SC-FETs in Fig.
\ref{fig:FCTransfer} are offset by 39~V while basically
maintaining the shape of the subthreshold region. Similar results
for SC-FETs have recently been reported by Takeya et al.
\cite{Takeya04}. Also in Fig. \ref{fig:FCTransfer}, we show the
transfer characteristic of a typical thin film device with
treatment (E). The curve for the thin film transistor was
normalized to account for the different W/L ratios. A detailed
analysis of the subthreshold region shows a very steep
subthreshold swing of 0.3~V/dec (300~nm SiO$_2$) for the SC-FET
treated with (E), and a slightly larger subthreshold swing of
0.5~V/dec for the SC-FET with (A). The significantly steeper
subthreshold swing of the single crystal devices compared to that
of the thin film devices indicates a significantly lower trap
density for single crystal FETs.

The off-current in both single crystal devices is similar,
while it is an order of magnitude higher in the thin film
device, indicating that the off-current in the TFT could be
limited by bulk traps \cite{Voelkel02}\cite{Oberhoff04}.

While the behavior of the thin film devices can be modeled
using a flexible density of states model as shown in
\cite{Voelkel02}\cite{Oberhoff04}, modeling the single
crystal devices proves difficult. However, shifting the
transfer characteristic by $\Delta V_g\approx 40$~V while
maintaining a steep subthreshold swing is difficult to
achieve in a "trap-only" model, since the the amount of
acceptor states necessary to shift the turn-on voltage also
degrades the subthreshold swing. We take this as compelling
evidence that the observed shifts of the threshold voltage
and the turn-on voltage are caused by the built-in electric
field of the self assembled monolayers.

\section{Conclusions}
We fabricated pentacene thin film transistors and rubrene
single crystal FETs incorporating nine organosilanes with
different functional groups. The organosilanes form self
assembled monolayers on the SiO$_2$ gate insulator and have
various dipole moments depending on the electron acceptance
properties of their functional group. We find the dipole
moment of the SAM modifies the surface potential of the
layer next to the transistor channel and induces mobile
charge carriers at zero gate bias. This manifests itself in
a shift of the transfer characteristics. A simple energy
level diagram is used to explain these observations.
Similar shifts have been modeled by V\"{o}lkel et al.
\cite{Voelkel02} using appropriate trap state
distributions. From the difference between the threshold
voltage and the turn-on voltage we estimate the trap
density in the thin film FETs to be of order $1\times
10^{12}$/cm$^2$, while a lower trap density is found for
rubrene single crystal FETs. The single crystal experiments
clearly show that the built-in electric field of a self
assembled monolayer next to the transistor channel acts as
a gate bias and modulates the charge carrier density.

\begin{acknowledgments}
The authors would like to thank Kurt Mattenberger and
Hans-Peter Staub for technical solutions, Cornelius
Krellner for crystal growth and A. Stemmer for providing
access to the AFM equipment. Fruitful discussions with
Benjamin R\"{o}ssner, H. B\"{a}ssler, G. Horowitz, T. N. Jackson,
E. J. Meijer, G. Paasch, S. Scheinert and K. Seki are
gratefully acknowledged. Further we would like to thank Urs
Notter and the other members of the machine shop of the ETH
Physics Departement. This study is partly supported by ETH
grant 20020-02, by the Swiss National Science Foundation,
and by the Swiss BBW as part of the EU-Research program
EUROFET (HPRN-CT-2002-00327).
\end{acknowledgments}

\newpage

\bibliographystyle{apsrev}
\bibliography{../../../TeX/References}

\end{document}